\documentclass[11pt]{article} 

\usepackage[utf8]{inputenc} 

\usepackage{geometry} 
\usepackage{graphicx} 
\usepackage{flafter} 
\usepackage{sidecap} 
\usepackage{subfig} 

\usepackage[font=small, labelfont=bf, format=hang]{caption}

\usepackage{tabularx}
\usepackage{booktabs} 
\usepackage{array} 
\usepackage{paralist} 
\usepackage{verbatim} 
\usepackage{subfig} 

\usepackage{fancyhdr} 
\usepackage{sectsty}
\allsectionsfont{\sffamily\mdseries\upshape} 

\usepackage[nottoc,notlof,notlot]{tocbibind} 
\usepackage[titles,subfigure]{tocloft} 

\usepackage{authblk}

\title{Neural Hypernetwork Approach for Pulmonary Embolism diagnosis}
\author[1]{Matteo Rucco\thanks{matteo.rucco@unicam.it}}
\author[2]{David M. S. Rodrigues\thanks{david.rodrigues@open.ac.uk}}
\author[1]{Emanuela Merelli}
\author[2]{Jeffrey H. Johnson}
\author[3]{Lorenzo Falsetti}
\author[3]{Cinzia Nitti}
\author[3]{Aldo Salvi}

\affil[1]{University of Camerino, School of Science and Technology}
\affil[2]{Design Complexity Group, The Open University, United Kingdom}
\affil[3]{Internal and Subintensive Medicine of Ospedali Riuniti - Ancona, IT}

\begin{document}
  \maketitle
\section{Abstract}

The purpose of the work is to introduce an integrative approach for the analysis of partial and incomplete datasets that is based on Q-analysis with machine learning. The new approach, called Neural Hypernetwork, has been applied to a case study of pulmonary embolism diagnosis. The objective of the application of neural hypernetwork to pulmonary embolism (PE) is to improve diagnose, henceforth helping in reducing the number of CT-angiography scans needed for confirmation of the condition. Hypernetworks are based on topological simplicial complexes and generalize the concept of two-body relation to many-body relation. Furthermore, Hypernetworks provide a significant generalisation of network theory, enabling the integration of relational structure, logic and analytic dynamics. Another important results is that Q-analysis stays close to the data, while other approaches manipulate data, projecting them into metric spaces or applying some filtering functions to highlight the intrinsic relations. A pulmonary embolism (PE) is a blockage of the main artery of the lung or one of its branches, frequently fatal. Our study uses data on 28 diagnostic features of 1,427 people considered to be at risk of PE. The resulting neural hypernetwork correctly recognised 94\% of those developing Pulmonary Embolism. This is better than previous results obtained with other methods (statistical selection of features, partial least squares regression, topological data analysis in a metric space).\\\\
\textit{\textbf{Keywords}: TOPDRIM, Hypernetworks, Q-analysis, Pulmonary Embolism, Topology~of~data}

\section{Introduction}
Pulmonary embolism (PE) is still a disease without a set of specific clinical variables (CV). During the diagnostic stage medical doctors (MDs) collect a large dataset, in the order of 30 CVs, to describe the health status of a patient.  Clinical Prediction Rules (CPR) are then computed using a subset of these CVs and are used to assign to each patient a probability of being sick. The most used CPRs in diagnosing PE are Wells and Revised Geneva scores. Nevertheless, in most cases MDs still need to perform a CT-angiography to obtain a final conclusive diagnose. CT-angiography is still a dangerous test, due to the ionization radiation and with high costs for medical structures. At the best of our knowledge only few papers regarding advanced methods for improving detection of pulmonary embolism (PE) have been published. For the sake of clarity, two of these have been written by Tourassi et al., in which they have analysed the Prospective Investigation of Pulmonary Embolism Diagnosis (PIOPED) database; in the third paper by Rucco et al., the authors commented the results obtained with the combined application of topological data analysis in a metric space and artificial neural network. In all these papers the performance of the classifiers are good enough to suggest that a Computer Aided Detection (CAD) for PE is reasonable, however the results of the third paper can not be directly compared with the result of the two others because they used a different cohorts. Here we propose a new CAD based on the mathematical theory of Hypernetwork~\cite{D13} with Q-analysis~\cite{D3,D6,D13,D14} for a deep study of patients dataset and then we used the results for the feature selection for the training step of an artificial neural network.  The dataset used in this work is the same used by Rucco et al. Our study uses data on 28 diagnostic features of 1,427 people considered to be at risk of PE and the outcome for each person of whether or not the developed a PE. Our approach correctly recognised 94\% of those developing a PE.

\subsection{Hypernetwork}
Hypernetwork theory~\cite{D3, D10, D11, D13, D15} concerns the formation of combinations of entities under relationships to form Hypersimplices. For example, the combination of attributes $\langle elderly, poor sight, live alone\rangle $ may make a person more predisposed to a fall than, say, $\langle young, normal sight, cohabit\rangle$ . The entities in the brackets are called vertices since hypersimplices generalise networks. Two vertices $\langle a, b\rangle$  correspond to the usual edge in a network, three vertices correspond to a triangle $\langle a, b, c\rangle$ , four vertices correspond to a tetrahedron $\langle a, b, c, d\rangle$ , and so on. From this perspective it is natural to see the 28 features defined above as a hypersimplex with 28 vertices, $\langle  f_{1}, f_{2}, … , f_{28}; R \rangle$ . The symbols R is the relation that binds the vertices together. Hypernetwork theory makes a distinction between a vertex with a measurement on it, and a vertex as a class of values associated with a scale. This also provides a coherent way of handling missing data.  In the simplest case an observed variable may just be a categorical variable stating that a condition is either present or not-present. Rather than treat this as a single vertex with a number on it, 0 meaning not-present and 1 meaning present, hypernetwork theory treats this situation by having two vertices, ‘x is not present’ and ‘x is present’. Then for the case of missing data neither of the vertices is related to the patient. Not that ‘observed as not being present’ is very different from ‘not being observed’, and this affects subsequent analyses. In other cases the interpretation of data as a number associated with a vertex on a ratio scale may not reflect the way the data were collected and what they mean. Very often the numbers on the scale are ordinal and should be interpreted, for example, as low, medium and high classes defined by three vertices in this case: x-low, x-medium, and x-high. This is effectively a discretisation of the scale. In hypernetwork terms this makes very good sense. Consider four pairs of diagnostic vertices $a, \sim a, b, \sim b, c, \sim c, d, \sim d$, where $\sim x$ means ‘not x’. In hypernetwork theory $\sim x$ is called the antivertex of x. When all the data are available every subject will be represented as a tetrahedron such as $\langle a, b, \sim c, d\rangle$ . Suppose that the data for d or $\sim d$ were recorded a NaN, with no information for either vertex. Then the simplex as $\langle a, b, \sim c\rangle$  represents that part of diagnostic space for which there is information. The simplex $\langle a, b, \sim c\rangle$ , which  is a triangle and a face of the tetrahedra $\langle a, b, \sim c, d\rangle$  and $\langle a, b, \sim c, \sim d\rangle$ , is the best representation of the available data.

\subsection{Artificial Neural Network}
An artificial neural network is a computational metaphor inspired by the biological brain’s network. The neural network used in our study has a three-layer, feed-forward architecture and was trained by using the back-propagation algorithm with a sigmoid activation function. According to this learning scheme, the network tries to adjust its weights so that for every training input it can produce the desired output. A supervised learning strategy was used where during the training phase, the network is presented with pairs of input-output patterns. It has been shown that this technique minimises the mean squared error (MSE) between the desired and the actual network output following an iterative gradient search technique. The number of hidden layer neurones was determined experimentally evaluating trial and error, as there is no theory that allows the prediction the correct number of neurones for the best topology of the network that ensures the optimal sensitivity and specificity~\cite{1, 2}.

\subsection{The Pulmonary Embolism Case Study}
PE is still difficult to diagnose because clinical symptoms and signs are nonspecific. Among the patients who die of PE, the majority of the deaths is observed during the first few hours after the acute event~\cite{6}. Despite diagnostic advances, delays in pulmonary embolism diagnosis are common and represent an important issue~\cite{7}. As a cause of sudden death, massive PE is second only to arrhythmic death. Among survivors, recurrent embolism and death can be prevented with prompt diagnosis and therapy. Unfortunately, diagnosis is often missed because patients often show nonspecific signs and symptoms. If left untreated, approximately one third of patients who survive an initial pulmonary embolism die from a subsequent embolic episode ~\cite{8}. Independently to the anatomic extension of the embolism, the vascular involvement and the dimensions of the thrombus, it is now widely accepted that the most important prognostic marker is hemodynamic compromise. Classically, PE has been subdivided in massive, hemodynamically unstable (shock, hypotension or cardiac arrest), sub-massive (normotensive with right ventricle dysfunction) or non-massive (normal blood pressure and no signs of right ventricle dysfunction). The highest mortality is observed the first two categories. Shock, right ventricle dilatation at echocardiography and myocardial damage as assessed by Troponin I levels are deemed to be the most important and widely recognised elements that can predict adverse outcomes in PE. The role of "newer" markers, such as BNP, even if included in some PE prognostic models ~\cite{9} and associated to a worse prognosis, is less clear. In our opinion, a significative increase of BNP levels can be interpreted as an global expression of a right ventricle dysfunction (RVD). These markers maintain a predictive role also among patients presenting with normal blood pressure and without signs of shock: RVD, defined as the presence of increased troponin values or echocardiographic evidence of right ventricle dilatation or increased pulmonary pressure, has been associated to a the higher mortality ~\cite{10} in all the subset of patients.

\section{Methodology}
\subsection{Missing data and Discretisation in Machine Learning}
Discretisation of continuous features plays an important role in machine learning techniques either because the machine learning technique itself requires a nominal feature space or because discretisation allows better results of the operating machine learning technique. The field of research on dataset discretisation for machine learning is vast and beyond the scope of this paper, but it is important to say that such algorithms usually aim to maximize the interdependency between discrete attribute values and class labels, as this minimizes the information loss due to the discretisation process. The process of discretisation has to balance the trade off between these two goals and many studies have shown that several machine-learning techniques benefit from this discretisation process ~\cite{15, 16, 17}.  In the present study, the transformation process of the dataset by discretisation and duplication of the features variables aims also to improve quality of the machine learning technique in dealing with missing data. 

In the present study, the dataset is very noise with 24\% of missing data. This forces the use of some data transformation technique that allows the use of the entire dataset with machine learning techniques. Dealing with missing data presents a challenge and many techniques can be used to overcome these difficulties.  Typically missing data can be categorised according to the statistical properties of the occurrences of the missing data. Either the missing data is missing completely at random (MCAR, independent of that variable y or other variables x), is missing at random (MAR, missing data depends on other variables x), or it is missing not at random (MNAR, dependence of self variable and others).  Several techniques can be applied to deal with the different types of missing data and the most common are deletion methods, imputation methods and model based methods. Deletion methods remove the missing data from the dataset either in a listwise manner or a pairwise manner. This kind of procedure introduces a bias in the resulting dataset if the missing data is of type MNAR. Imputation methods generate values for missing data from statistical measures like sample mean and mode and are therefore difficult to apply to class variables. Also they reduce variability and weaken covariance and correlation estimates in the data because they ignore relationships between variables. Model–based usually use maximum likelihood estimators or multiple imputations. Although dealing with the entire dataset, the estimated dataset present problems of bias in the case of MNAR. Taking this into consideration we developed a mechanism of dealing with the missing data in the dataset. The dataset is naturally MNAR as the missing data is highly dependent on the observed variable as it is also dependent on other variables. This means that deletion methods will reduce (drastically) the number of observations and will introduce bias. The imputation and model-based methods are also of difficult application due to the existence of nominal categories in the dataset.  By the combination of discretisation (needed to improve machine learning techniques) we were able to solve also the problem of the missing data in the dataset. 

The discretisation was done manually dependent on medical indications of what would consist health risk for each factor studied. For each factor a pair of boolean variables were created that represent if the observed factor “doesn’t contribute to an health risk” and if the observed factor “contributes to an health risk”. In terms of \emph{Q}-analysis, this corresponds to the transformation of each variable in a vertex, and its antivertex~\cite{D13}. This technique allows information to be kept about missing data by setting missing data entries into those pairs of vertices as $\langle false,false\rangle$ ($\langle0,0\rangle$ for computation purposes). On the other hand if there is an observation for that factor it will be represented in these pairs of vertices $\langle false,true\rangle$ or $\langle true,false\rangle$($\langle0,1\rangle$ or $\langle 1,0 \rangle$ for computational purposes). These factors are discretised by taking into account the risk ranges discussed with the medical staff and therefore represent a data driven solution for the discretisation problem. In this way one doesn’t incur in the problems of traditional discretisation techniques. The resulting dataset includes more information about the original dataset then a deletion technique would produce and therefore the subsequent machine learning technique benefits from the extended dataset set used during the learning phase of the algorithm. 

\subsection{Dataset description}
The dataset of the pulmonary embolism is constituted by 1430 samples corresponding each entry to a patient and 28 variables, where 26 of these variables are clinical indicators. One variable corresponds to the patient ID and one variable indicates the final disease diagnostic on the pulmonary embolism condition.

\begin{table}[htpb!]
\caption{The original 28 variables}
\label{tb:a}
\begin{tabularx}{\textwidth}{|l|r|X|}
\hline
	1 & ID & Patient's identifier     \\ \hline
	2 & Age & With the increase of the age, increase the incidence  \\ \hline
	3 & N\_F\_Pred & Number of predictive factors \\ \hline
	4 & N\_F\_Risk & Number of risk factors \\ \hline
	5 & Previous DVT & A previous DVT / PE is a risk factor repeated infringement DVT / PE \\ \hline
	6 & Palpitations & Aspecific symptom. If it implies a tachycardia could be associated with DVT/PE \\ \hline
	7 & Cough & Coughs Symptom very nonspecific but frequently present in patients with DVT / PE \\ \hline
	8 & dDimer & A value of d-Dimer < 230ng=ml is associated with a low risk / absent of DVT/ PE. A very high value is associated with a high risk of DVT / PE \\ \hline
	9 & PAS & A low PAS is present in patients with DVT / PE and hemodynamic shock \\ \hline
	10 & PAD & In cardiogenic shock with DVT / PE is low, sometimes undetectable. By itself has no value despite of the PAS \\ \hline
	11 & FC & In the patient with TVP / EP tachycardia is often found \\ \hline
	12 & PAPS & It is one of the criteria of right ventricular dysfunction. It can be normal in the case of EP low entity. \\ \hline
	13 & WBC & The value increases with inflammatory forms (pneumonia, etc ...) that can be confused with DVT / PE  \\ \hline
	14 & Cancer at diagnosis & Cancer at diagnosis It is a risk factor for DVT / EP recognized \\ \hline
	15 & Troponin & It is a marker of myocardial infarction or heart failure and can be confused with DVT / PE \\ \hline
	16 & Shockindex & It is the ratio between PAS and FC, if it is greater than 1 is indicative of shock \\ \hline
	17 & NEOPLASIA & It is a risk factor for DVT / EP recognized \\ \hline
	18 & RVD & Right ventricular overload in the course of DVT / PE \\ \hline
	19 & Wells score & Wells score     \\ \hline
	20 & Revised Geneva score & Revised Geneva   \\ \hline
	21 & Wicki score & Wicki’ score  \\ \hline
	22 & Dyspnea & Main symptom in DVT / PE \\ \hline
	23 & Chest pain & Chest pain is present in myocardial infarction, in pleural effusion, in the high DVT \\ \hline
	24 & PCO2 & Associated with low pO2 may be suggestive of DVT / PE  \\ \hline
	25 & PO2 & Associated with low pCO2 may be suggestive of DVT / PE \\ \hline
	26 & PH & In DVT / EP pH is usually normal \\ \hline
	27 & Hemoptysis & It is the expectoration (coughing up) of blood or of blood-stained sputum from the bronchi, larynx, trachea, or lungs \\ \hline
	28 & Final diagnose & Final physicians' diagnosis \\ \hline
\end{tabularx}
\end{table}

The original dataset includes approximately 24\% of empty elements (NaN). These are elements for which a value is unknown either because the patient wasn’t asked or because they did not provide an answer. The distribution of the missing data is not uniform and in the case of some variables they represent 1227 out of the 1430 observations (86\%). A detailed breakdown of how the different variables are affected by this can be found in table~\ref{tb:b}.

\begin{table}[htpb!]
\caption{NaNs percentage}
\label{tb:b}
\begin{tabularx}{\textwidth}{|X|X|X|}

\hline
	Feature & NaNs & \% of NaNs \\ \hline
	ID & 0 & 0.00 \\ \hline
	Age\_CAL & 1 & 0.07 \\ \hline
	N\_F\_PRED & 95 & 6.64 \\ \hline
	N\_F\_Risk & 108 & 7.55 \\ \hline
	Previous DVT & 490 & 34.27 \\ \hline
	Palpitations & 196 & 13.71 \\ \hline
	Cough & 184 & 12.87 \\ \hline
	dDimer & 111 & 7.76 \\ \hline
	PAS & 162 & 11.33 \\ \hline
	PAD & 546 & 38.18 \\ \hline
	FC & 58 & 4.06 \\ \hline
	PAPS & 1227 & 85.80 \\ \hline
	WBC & 38 & 2.66 \\ \hline
	Cancer at diagnosis & 670 & 46.85 \\ \hline
	Troponin & 571 & 39.93 \\ \hline
	Shockindex & 170 & 11.89 \\ \hline
	Cancer & 668 & 46.71 \\ \hline
	RVD & 862 & 60.28 \\ \hline
	Wells score & 512 & 35.80 \\ \hline
	Revised Geneva score & 0 & 0.00 \\ \hline
	Wicki score & 665 & 46.50 \\ \hline
	Dyspnea & 186 & 13.01 \\ \hline
	Chest pain & 184 & 12.87 \\ \hline
	PCO2 & 242 & 16.92 \\ \hline
	PO2 & 238 & 16.64 \\ \hline
	PH & 577 & 40.35 \\ \hline
	Hemoptysis & 490 & 34.27 \\ \hline
	Final Diagnosis  & 0 & 0.00 \\ \hline
	TOTAL & 9251 & 23.10 \\ \hline
\end{tabularx}
\end{table}

From the original dataset a subset of 19 clinical variables has been selected. We discarded the following: 
\begin{itemize}

\item ID has been used only to identify uniquely each patient;
\item Troponin, PAS, PAD and PAPS because they represent information already deduced from other clinical variables or due to the huge amount of missing data (PAPS);
\item Wells, Revised Geneva and Wiki score because they have been thought as the synthesis of other clinical variables plus the MDs opinion and they represent the clinical prediction rules actually used in the hospital.
\item Final diagnosis has been used during for labelling the input for the artificial neural network
\end{itemize}

The application of the thresholding procedure projected the 19 clinical variables in 38 descriptors:

\begin{table}[ht!]
\caption{Descriptors used in the neural network experiment.}
\label{tb:c}
\begin{tabularx}{\textwidth}{|X|X|}
\hline
	1. Age [..., 64]  & 2. Age [65; ...] \\ \hline
	3. N\_F\_Pred [0] & 4. N\_F\_Pred [1, ...]  \\ \hline
	5. N\_F\_Risk  [0] & 6. N\_F\_Risk [ 1, ...] \\ \hline
	7. Previous DVT [0] & 8. Previous DVT 0 [1] \\ \hline
	9. Palpitations [0] & 10. Palpitations [1] \\ \hline
	11.Cough [0] & 12.Cough [1] \\ \hline
	13. dDimer  [<=230] & 14. dDimer  [>230] \\ \hline
	15. FC [50 - 99]  & 16. FC [100, ...] \\ \hline
	17. WBC [2000; -10000]  & 18. WBC [10000, ... ] \\ \hline
	19. Cancer at diagnosis [0] & 20. Cancer at diagnosis [1] \\ \hline
	21. Shockindex [... 0:89]  & 22. Shockindex [0:9; ...] \\ \hline
	23. Cancer [0]  & 24. Cancer [1] \\ \hline
	25. RVD [0] & 26. RVD [1] \\ \hline
	27. Dyspnea [0] & 28. Dyspnea [1] \\ \hline
	29. Chest pain [0] & 30. Chest pain [1] \\ \hline
	31. PCO2 [35- 45] & 32. PCO2 [35  45] \\ \hline
	33. PO2 [ - 60] & 34. PO2 ]60, …] \\ \hline
	35. PH [7:3 - 7:42] & 36. PH [7:42, …] \\ \hline
	37. Hemoptysis [0] & 38. Hemoptysis [1] \\ \hline
	39. Final diagnosis [0] & 39. Final diagnosis [1] \\ \hline
\end{tabularx}
\end{table}

\section{Results}

Q-analysis allows for a description of the dataset in terms of the relations existing in it between patients and variables. This technique allows the identification of the structural properties (the backcloth) of the dataset in terms of describing the dimensionality of the connectivity of the variables in terms of the number of patients that share those variables ~\cite{D12, D13}. Naturally it is important to describe the dataset in terms of what happens to the variable representing the final diagnose of the patients (\textit{Final diagnosis [0]} for negative diagnosis and \textit{Final diagnosis [1]} for positive diagnosis.).

The positive diagnosis \textit{Final diagnosis [1]} variable occurs 819 times while the \textit{Final diagnosis [0]} occurs only 608 times. This means that they can only be q-connected up to dimensions of 818 and 607 respectively. It is therefore interesting to identify what happens to these two variables. Both are present in the same q-connected component up to the dimension q=529, when \textit{Final diagnosis [0]} becomes an isolated variable and \textit{Final diagnosis [1]} persists in a connected component with another 11 variables. This 12 element component persists until \textit{Final diagnosis [0]} disappears at $q>607$ and beyond until q=707, where the component starts to slowly get disconnected. At q=720 \textit{Final diagnosis [1]} is still connected to 9 variables and it becomes isolated by q=721. It stays in isolated from that dimension until q=817 when it disappears. 

\begin{figure}[ht!]
    \centering
    \includegraphics[width=0.8\textwidth]{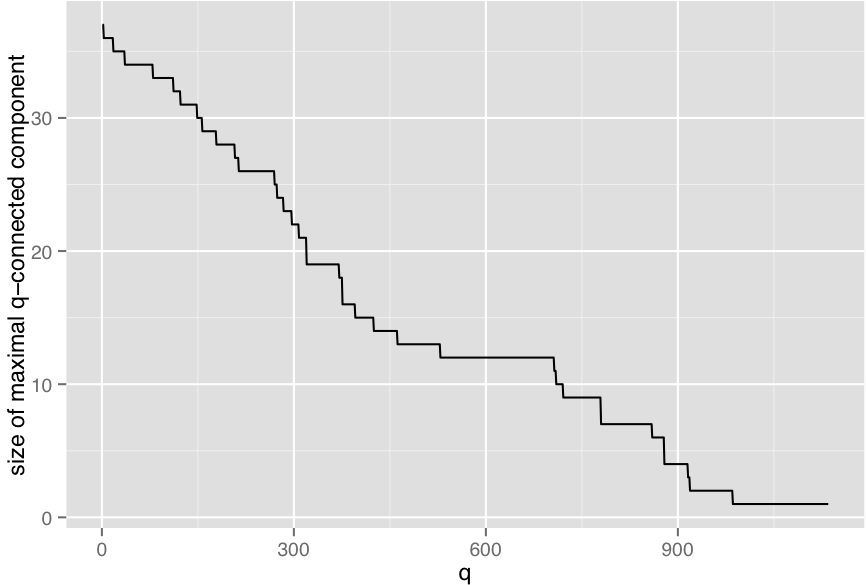}
    \caption{Size of maximal connected component as a function of the dimension of the q}
    \label{fig:connected}
\end{figure}
The identification of that of the stability zone can be clearly observed in the Figure~\ref{fig:connected} where the maximal size connected component is plotted as a function of the dimension of the induced q-graph. 

The 12 factors identified in plateau in the q-range=[529,707] the \textit{Final diagnosis [0]}  connected component are \textit{age [1]} , \textit{cough [0]} , \textit{shockindex [0]} ,\textit{PO2 [1]}, \textit{previous DVT [0]}, \textit{palpitations [0]}, \textit{dDimer [1]} , \textit{FC [1]} , \textit{dyspnea [1]}, \textit{chest pain [0]} , \textit{hemoptysis [0]}, and naturally \textit{final diagnosis [1]} .

This backcloth reveals the connectivity of the variables at this dimension, but it isn’t an account of the causal or even correlation effect of those variables with the diagnosis. To infer these we used a neural network approach using as input all the existing variables.

In any case, Q-analysis reveals that this high dimensional component of the variables is subset that is shared among many patients making them general and non-discriminative in terms of diagnose, for short we call this set “connected higher dimensional set” We tested this by training an ANN with these features and it result in AUC of 60\%. Variables that are not connected (unconnected higher dimensional level set) in an higher dimensional level are therefore very important for inclusion in the ANN training. Training the ANN with the complementary set of variables yielded an AUC of 86\%. This confirms that the diagnose can’t be made by the structural backcloth of the relation between the variables, but by the traffic existing in the structure. 

\begin{figure}[ht!]
    \centering
    \includegraphics[width=0.8\textwidth]{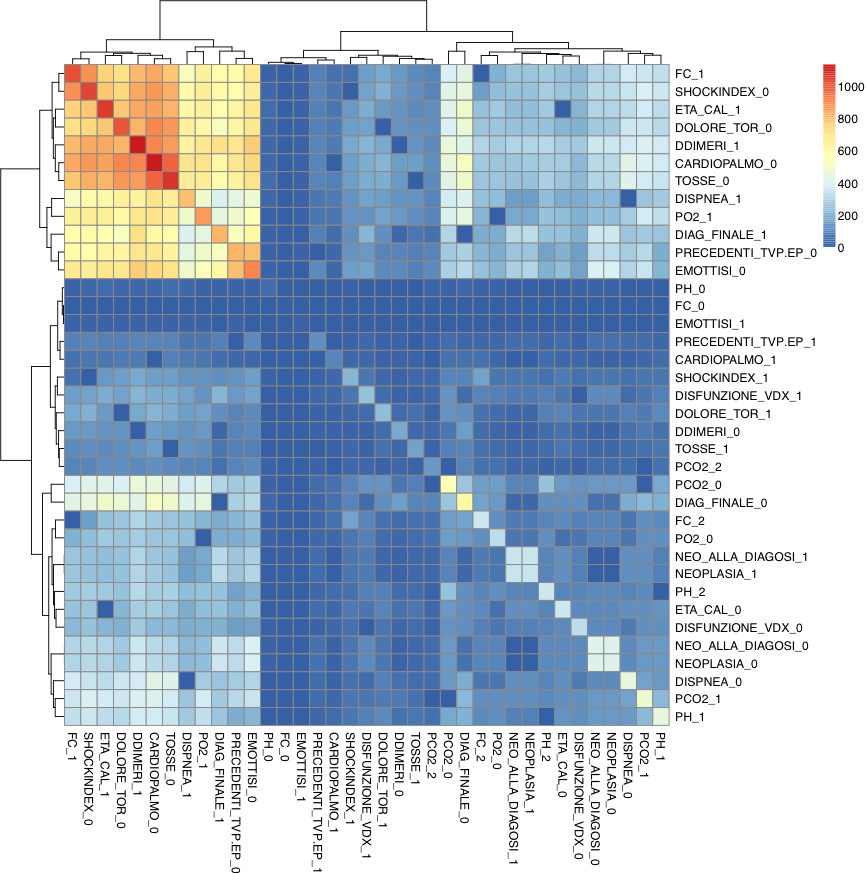}
    \caption{Clustering of the Shared Face Matrix of the Pulmonary Embolism variables}
    \label{fig:heat}
\end{figure}

This result is clearly observed also by clustering the element of the shared face matrix with an Euclidean distance as in figure \ref{fig:heat}. It is clear from the clustering that the top-left block is highly correlated and share the highest number of patients. 

\begin{figure}[ht!]
    \centering
    \includegraphics[width=0.8\textwidth]{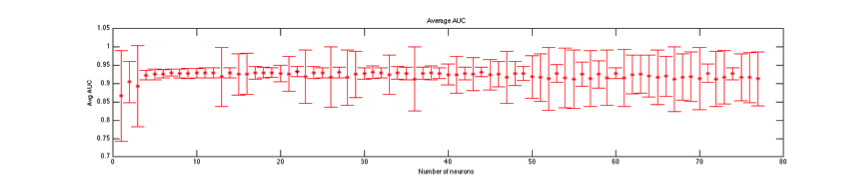}
    \caption{Study of number of neurons: y-axis average AUCs, x-axis number of neurons. The blue point represents the optimal configuration}
    \label{fig:neurons}
\end{figure}

\begin{figure}[ht!]
    \centering
    \includegraphics[width=0.8\textwidth]{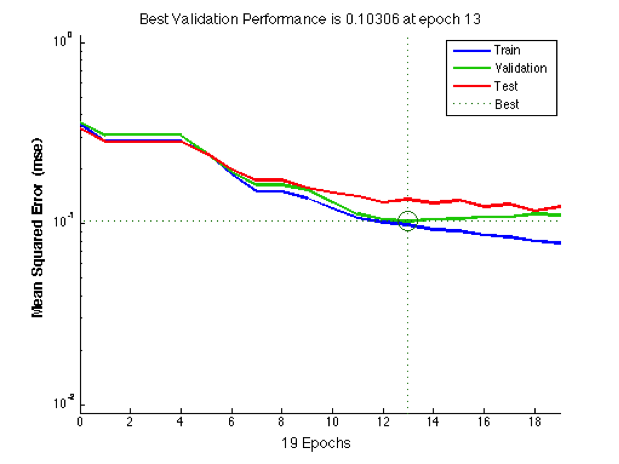}
    \caption{Best Mean Squared Error (MSE) at 13 epochs}
    \label{fig:error}
\end{figure}

\begin{figure}[ht!]
    \centering
    \includegraphics[width=0.8\textwidth]{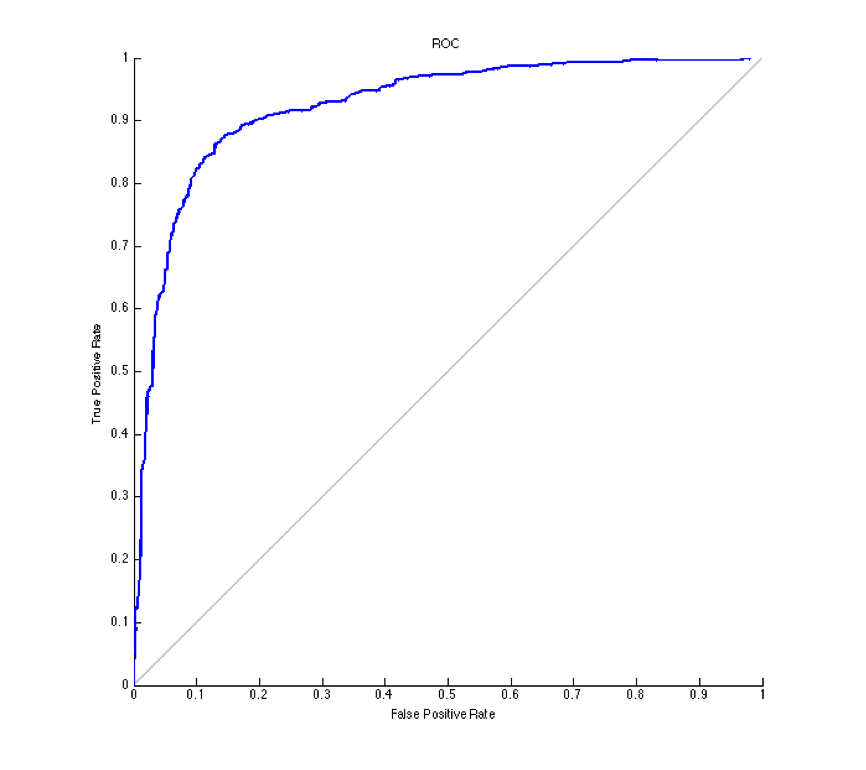}
    \caption{ROC curve for the new classifier based on Artificial Neural Network with AUC = 93\%}
    \label{fig:auc}
\end{figure}

At last stage, we trained an artificial neural network using the full thresholded dataset. The input of the network is a subset formed with the 70\% randomly selected patients from their dataset presented to the network with the features used in the Q-Analysis. To define the optimal architecture of the ANN we studied 76 combinations of neurons in the hidden layer, from 1 neuron to 77, and for all these we executed 100 trials with the permutations of rows in the patients' dataset, at the end we designed and coded in Matlab ~\cite{3} a network with 22 neurons in the hidden layer (\ref{fig:neurons}), and an output layer with a single decision node. The learning rate was selected to be 0.5 and the momentum coefficient to be 0.9, the optimal number of iterations (epochs) have been found equal to 13  and it corresponds to a Mean Squared Error (MSE) equal to 0.10306 (see \ref{fig:error}). The AUC of the ROC curve (see figure \ref{fig:auc}) is equal to 93\%. In this study, the network's output was interpreted as the probability of PE being present.

\newpage
\section{Conclusion}

We introduced an novel integrative approach for the analysis of partial and incomplete datasets that is based on \emph{Q}-analysis coupled with machine learning. The new approach, called Neural Hypernetwork, has been applied to a case study of pulmonary embolism diagnosis with great success as shown by the 93\% of the AUC of the ROC (fig~\ref{fig:auc}).

The results of this study can be compared to the findings of previous works that have taken  different approaches. The comparison is summarised in table \ref{tb:d} and clearly shows that this novel approach supplants previous results. 

\begin{table}[ht!]
\caption{Comparison of performance}
\label{tb:d}
\begin{tabularx}{\textwidth}{ | X | X | X | X | X | X | X | X | X | }
\hline
	 & Wells + ANN & Rev. Geneva + ANN & Stat. feature reduction + ANN & Conn. higher dimen. set + ANN & PLS regression & Unconn. higher dimen. set + ANN & Ayasdi Iris + ANN & Neural Hypernetwork \\ \hline
	AUC (\%) & 0.74 & 0.55 & 0.45 & 0.6 & 0.74 & 0.86 & 0.89 & 0.93 \\ \hline
\end{tabularx}
\end{table}

In detail we studied the performance of the two scores: revised Geneva and Wells. The scores have been used independently to train an artificial neural network. An alternative approach for the features selection has been performed. We used the Mood’s variance test and the relative p-value to highlight the feature set with more discriminating power between the healthy and pathological classes. The authors have found results completely comparable with the literature (i.e. d-Dimer is the clinical variable that must be used as first clinical test in the pulmonary embolism diagnose procedure). With these extracted features we trained an ANN  and training of an ANN with the selected features ~\cite{18}. We applied also a more classical methods: Partial Least Square Regression ~\cite{19} the best performance of the method has been found using the entire set of components (18) due to the low variance expressed by each single variables. The number of components has been found iteratively: for each iteration the rows of the initial dataset have been shuffled randomly and both the Jaccard coefficient (between the fitted response and the expected response) and the AUC have been evaluated

\begin{figure}[ht!]
    \centering
    \includegraphics[width=0.8\textwidth]{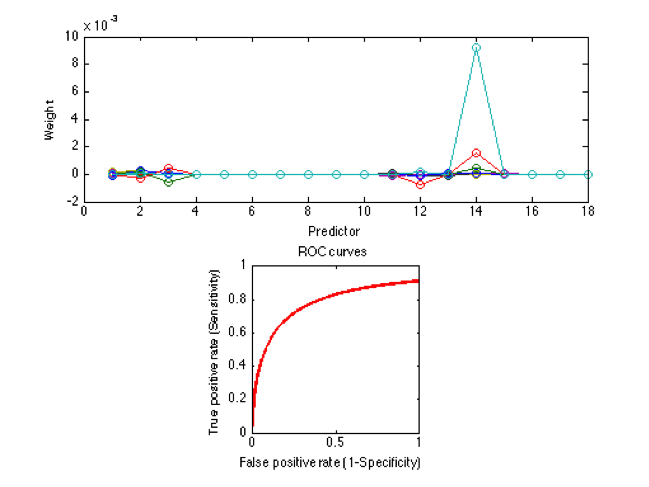}
    \caption{weight contributes in each component. Bottom: ROC curves for a PLS with 18 components}
    \label{fig:auc2}
\end{figure}

In the past, some of the authors have carried out a topological data analysis using Iris (a software developed by Ayasdi Inc.) in which they have found some interesting cluster in the healthy and pathological groups and this information has been used to train an ANN ~\cite{Rucco}. 
The purpose of this study was to introduce an innovative hybrid approach based on the hypernetworks’ theory and supervised artificial neural networks. This method has been used to design a new CAD for pulmonary embolism with the aim to reduce the number of CT-angiography analysis and ensuring a good efficiency of the diagnose. \\
We argued that our results can be thought in terms of the \textit{S[B]} paradigm. S[B] is an innovative paradigm presented by Merelli et. al, in \cite{merelli2013non, merelli2014topology} for modelling complex systems with a multi-level approach. The paradigm is based on two interacting levels: \textit{Structural} and \textit{Beahavioural}. The two levels are entagled by a feedback loop: S fixed a set of constraints for the observables, the observables can work regarding the constraints or evolve highlighting the emergence of new model details. Bearing in mind this scenario it is easy to imagine the \textit{artificial neural network} as the structural level, namely the global description of the system based on the interaction among all the clinical variables, and the \textit{hypernetworks} with their \textit{q-analysis} synthesis as the behavioural counterparts.

\newpage

 \bibliographystyle{plain} 
 \bibliography{pe}

\end{document}